\documentclass[10pt]{iopart}
\usepackage{iopams}  
\usepackage{graphicx}
\usepackage{color}
\usepackage{hyperref}
\usepackage{comment}
\usepackage[normalem]{ulem}
\newcommand{\pd}{\partial}
\newcommand{\bd}[1]{\boldsymbol{#1}}

\newcommand{\ie}{{i.e., }}
\newcommand{\mc}[1]{\mathcal{#1}}

\newcommand{\avg}[1]{\left\langle{#1}\right\rangle}
\newcommand{\exb}{{E\times B}}
\newcommand{\ExB}{{\bd{E}\times \bd{B}}}
\begin{document}

\title[]{Effects of collisional ion orbit loss on neoclassical tokamak radial electric fields}

\author{Hongxuan Zhu, T. Stoltzfus-Dueck, R. Hager, S. Ku, and C. S. Chang}
\address{Princeton Plasma Physics Laboratory, Princeton, NJ 08543, USA}
\begin{abstract}
Ion orbit loss is considered important for generating the radially inward electric field $E_r$ in a tokamak edge plasma. In particular, this effect is emphasized in diverted tokamaks with a magnetic X point. In neoclassical equilibria, Coulomb collisions can scatter ions onto loss orbits and generate a radially outward current, which in steady state is balanced by the radially inward current from viscosity. To quantitatively measure this loss-orbit current in an edge pedestal, an ion-orbit-flux diagnostic has been implemented in the axisymmetric version of the gyrokinetic particle-in-cell code XGC. As the first application of this diagnostic, a neoclassical DIII-D H-mode plasma is studied using gyrokinetic ions and adiabatic electrons. The validity of the diagnostic is demonstrated by studying the collisional relaxation of $E_r$ in the core. After this demonstration, the loss-orbit current is numerically measured in the edge pedestal in quasisteady state. In this plasma, it is found that the radial electric force on ions from $E_r$ approximately balances the ion radial pressure gradient in the edge pedestal, with the radial force from the plasma flow term being a minor component. The effect of orbit loss on $E_r$ is found to be only mild. How ion orbit loss will affect $E_r$ in the full-current ITER plasma pedestal is left as a subsequent study topic.
\end{abstract}

%
%
%
%
 
\ioptwocol
\section{Introduction}

In tokamaks, the ion orbits have finite excursion widths across magnetic flux surfaces, so the ions at the plasma edge can move across the last closed flux surface (LCFS) and enter the scrape-off-layer (SOL) region. While some ions will cross the LCFS again and return to the confined region, others may hit the wall. This effect is known as ion orbit loss, and is considered important for the modeling of plasma properties at the edge, such as the radial electric field $E_r$ and the toroidal rotation. Ion orbit loss has been emphasized in diverted tokamaks with a magnetic X point, in which the orbit length from the LCFS to the divertor plates changes with the toroidal-magnetic-field direction, and the modified process is referred to as X-point ion orbit loss \cite{Chang02,Shaing02,Ku04,Stacey11}. Ion orbit loss has been studied by many authors \cite{Chang02,Shaing02,Ku04,Stacey11,Itoh88,Shaing89,Connor00,Stoltzfus12,Battaglia14,Boedo16,Chang04,Chang17,Ku18,Brzozowski19}, either in limiter- or divertor-tokamak geometry.  It is sometimes assumed that the ion distribution function is empty to the lowest order in the loss-orbit velocity space but is Maxwellian in other parts of the velocity space \cite{deGrassie09,deGrassie15,Pan14}. However, this assumed distribution function is known to be inaccurate due to effects such as collisional and turbulent scattering into and out of the loss-orbit portion of the velocity space \cite{Seo14}. The X-point ion-orbit-loss theory emphasizes the difference between the ``forward-$\nabla B$'' configuration when the magnetic drift points towards the X point, and the ``backward-$\nabla B$'' configuration when the magnetic drift points away from the X point \cite{Chang02,Shaing02,Ku04}. Quantitative kinetic evaluation of the effect of ion orbit loss on $E_r$, including changes in its effect between the two configurations, are desirable. 

In this paper, we study the  quantitative effects of ion gyrocenter orbit loss on neoclassical $E_r$ using an ion gyrokinetic neoclassical code with adiabatic electrons, in which the the total ion radial current $J_r=J_{\rm pol} +J_{\rm gc}$ must be equal to zero. Here, $J_{\rm pol}$ is the ion polarization current associated with the time evolution of $E_r$, and $J_{\rm gc}$ is the ion gyrocenter current. In a steady-state edge plasma, both $J_{\rm pol}$ and $J_{\rm gc}$ should separately vanish; namely, $J_{\rm pol}\propto-\pd_t E_r= 0$ and  $J_{\rm gc}=J_{\rm vis}+J_{\rm loss}=0$.
Here, $J_{\rm vis}$ is the ion gyrocenter current induced by neoclassical viscosity and $J_{\rm loss}$ is the ion gyrocenter current induced by steady-state collisional scattering of gyrocenters onto the loss orbits. Absent orbit loss, $J_{\rm vis}(E_r)=0$ corresponds to the standard neoclassical solution $E_r=E_r^{\rm neo(0)}$ \cite{Hinton73,Hazeltine74,Hirshman78,Hirshman81}. When orbit loss becomes important, from the relation $J_{\rm vis}=-J_{\rm loss}$, one can evaluate how $J_{\rm loss}$ drives $E_r$ away from $E_r^{\rm neo(0)}$ \cite{Chang02,deGrassie15,Shaing92a,Shaing92b}. Further, by comparing $J_{\rm loss}$ between the forward- and backward-$\nabla B$ configurations, one can quantitatively study how ion orbit loss depends on the direction of the magnetic drift.

To measure ion gyrocenter orbit-loss current $J_{\rm loss}$, we have implemented a new numerical diagnostic in the gyrokinetic particle-in-cell code XGC \cite{XGC}. This diagnostic is based on the recently proposed ion-orbit-flux formulation \cite{Stoltzfus20,Stoltzfus21}, which allows us to measure the separate contributions to the ion orbit loss from different transport mechanisms and sources.  To focus on the neoclassical physics where $J_{\rm loss}$ is caused solely by Coulomb collisional transport, we report results from electrostatic simulations using the axisymmetric version of XGC (XGCa)  in this paper. We first demonstrate the validity of our diagnostic by studying the collisional relaxation of $E_r$ in the core. After this demonstration,  we numerically measure $J_{\rm loss}$ and $E_r$ at the edge for an H-mode plasma profile in a DIII-D geometry where the pedestal width is comparable with an ion banana-orbit width. For the given pedestal plasma profile and without considering neutral particles, the radial electric force on ions from $E_r$ is found to approximately balance the ion radial pressure gradient, confirming the strong radial diamagnetic property of collisional tokamak plasmas \cite{Hirshman78}. The effect of $J_{\rm loss}$ on edge $E_r$ is found to be only mild, driving $E_r$ away from $E_r^{\rm neo(0)}$ by a few percent. We emphasize that this conclusion only applies to the quasisteady state without the neutral particle effect. Ion orbit loss has been found to play a much more significant role during transient states, such as the development of the pedestal \cite{Chang02}, but such transient states are beyond the scope of this article.

The role of collisional ion orbit loss in a full-current ITER edge plasma, where ion neoclassical effects are much weaker than the present-day tokamaks due to the much smaller ion banana width (compared to the pedestal width), is important and is left for a subsequent study.

The rest of the paper is organized as follows. \Sref{setup} describes the setup for the XGCa simulations and the implementation of the ion-orbit-flux diagnostic. \Sref{core} studies the collisional relaxation of $E_r$ in the core. \Sref{edge} studies the effects of ion orbit loss on $E_r$ at the edge. Conclusions and discussion are given in \sref{conclusion}.
\section{XGCa simulation setup and ion-orbit-flux diagnostic implementation}
\label{setup}
\subsection{XGCa simulation setup}
\begin{figure}
\centering
\includegraphics[width=1\columnwidth]{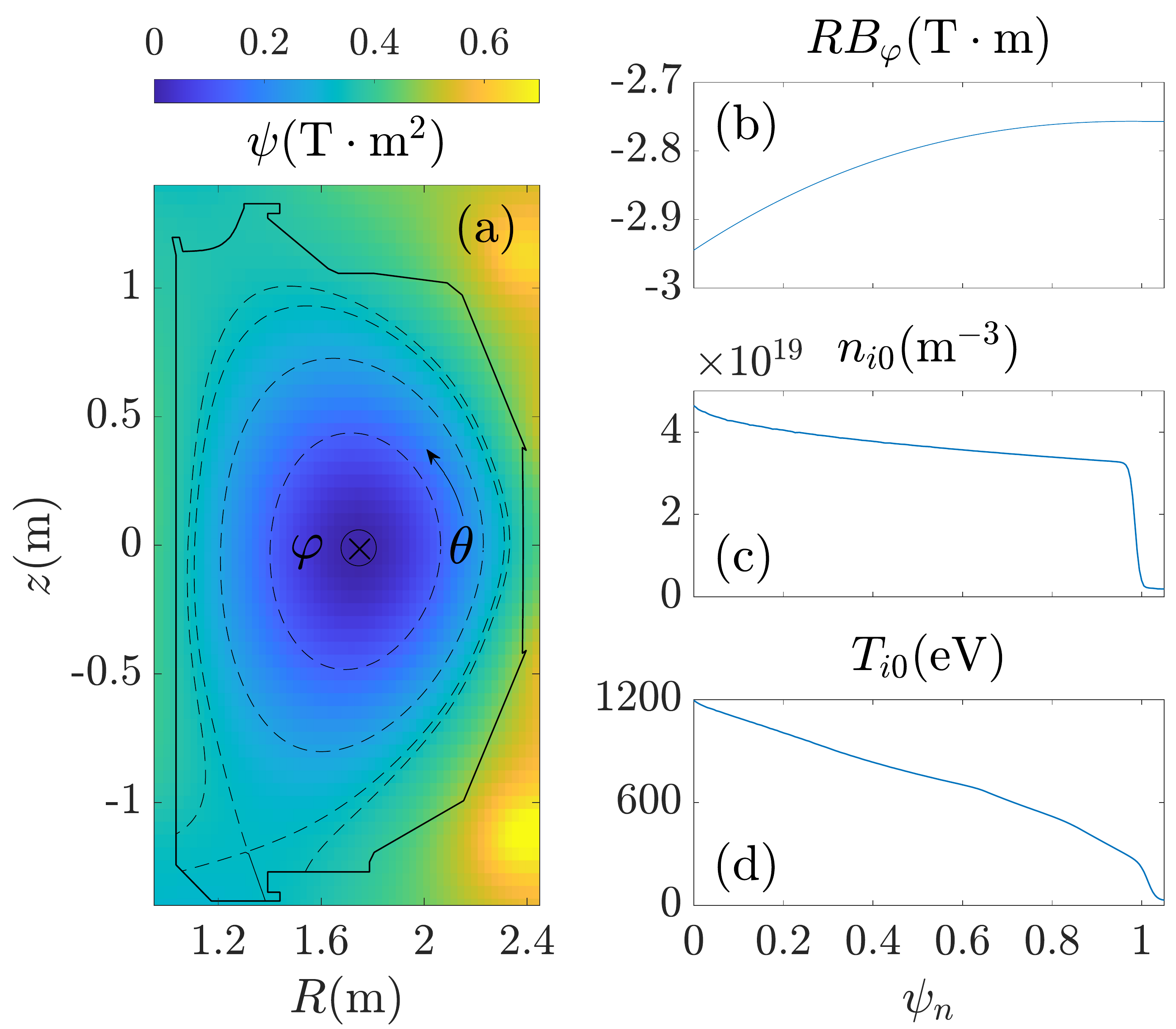}\caption{(a): The poloidal flux $\psi$ (shown by the color) from the DIII-D shot 141451 \cite{Muller11a,Muller11b,Seo14}.  The dashed curves show several flux surfaces and the solid curve is the wall.  Cylindrical coordinates $(R,\varphi,z)$ are used. We choose the poloidal angle $\theta$ to increase counterclockwise. (b-d): The equilibrium $I=RB_\varphi$, ion density $n_{i0}$, and ion temperature $T_{i0}$ versus the normalized flux  $\psi_n\doteq(\psi-\psi_{\rm a})/(\psi_X-\psi_{\rm a})$, where $\psi_{\rm a}$ and $\psi_X$ are the value of $\psi$ at the magnetic axis and the X point, respectively.}\label{equilibrium}
\end{figure}

We use electrostatic XGCa simulations to study an axisymmetric H-mode plasma in DIII-D geometry, with the plasma equilibrium from shot 141451 \cite{Seo14,Muller11a,Muller11b}  (\fref{equilibrium}).  The code uses cylindrical coordinates $\bd{R}\doteq(R,\varphi,z)$ to describe the realistic toroidal geometry containing an X point (\fref{equilibrium}(a)), where $\doteq$ means definitions.  The equilibrium magnetic field is given by
$\bd{B}=I(\psi)\nabla\varphi+\nabla\psi\times\nabla\varphi$,
where $\psi$ is the poloidal magnetic flux and $I=RB_\varphi$ is a flux function.  Since $B_\varphi<0$ (\fref{equilibrium}(b)),  the ion magnetic drift points in the negative-$z$ direction, which is the forward-$\nabla B$ configuration.  We will also simulate the backward-$\nabla B$ configuration where the sign of $B_\varphi$ is reversed ($B_\varphi>0$), while all other signs and plasma input profiles are kept the same. Since $\psi$ increases radially, the poloidal magnetic field $B_\theta$ is positive. If one reverses the sign of $B_\theta$ while keeping $B_\varphi$ unchanged, then the plasma will behave the same except that the toroidal-rotation direction is reversed.
The code utilizes unstructured triangular meshes, with most of the mesh nodes aligned with magnetic field lines \cite{Zhang16}. For our simulations, the mesh has a radial grid size of $\Delta\psi_n\approx 0.004$ and a poloidal grid size of $\Delta l_\theta\approx 1{\rm cm}$.  The normalized flux is defined as $\psi_n\doteq(\psi-\psi_{\rm a})/(\psi_X-\psi_{\rm a})$, where $\psi_{\rm a}$ and $\psi_X$ are the value of $\psi$ at the magnetic axis and at the LCFS, respectively.

We simulate deuterium ions with mass $m_i=2m_p$  and charge number $Z_i=1$,  where $m_p$ is the proton mass. The input ion density $n_{i0}(\psi)$ and temperature $T_{i0}(\psi)$ are functions only of  $\psi$, and are shown in figures~\ref{equilibrium}(c) and \ref{equilibrium}(d).  The ion gyrocenter coordinates are position $\bd{R}$, magnetic moment $\mu$, and parallel momentum $p_\parallel$. Their characteristics are governed by equations given in Ref.~\cite{Chang04}, which are mathematically equivalent to the following: 
\begin{eqnarray}
\label{XGC_Rdot}
B_\parallel^*\dot{\bd{R}}=(Z_ie)^{-1}\hat{\bd{b}}\times\nabla H+v_\parallel \bd{B}^*,\\
B_\parallel^* \dot{p}_\parallel=-\bd{B}^*\cdot\nabla H,
\end{eqnarray}
and $\dot{\mu}=0$. Here, the overhead dot denotes the time derivative, $\hat{\bd{b}}\doteq\bd{B}/B$, $\bd{B}^* \doteq\bd{B}+\nabla\times(p_\parallel\hat{\bd{b}}/Z_i e)$, $B_\parallel^*\doteq\hat{\bd{b}}\cdot\bd{B}^*$,  $H=p_\parallel^2/2m_i+\mu B+Z_ie \hat{J}_0{\Phi}$
is the gyrocenter Hamiltonian, $v_\parallel\doteq\pd_{p_\parallel}H$ is the parallel velocity, $e$ is the elementary charge,  $\hat{J}_0$ is the gyroaveraging operator, and $\Phi$ is the electrostatic potential.  Note that $v_\parallel=p_\parallel/m_i$ for the present electrostatic formulations. 

Using the  the ``total-f'' simulation method \cite{Ku16}, the code calculates both the total gyrocenter ion orbits and distribution function $F_i(\bd{R},\mu,p_\parallel,t)$. The gyrocenter distribution function is chosen to be Maxwellian at the beginning of the simulation:
\begin{equation}
F_i|_{t=0}=n_{i0}\left(\frac{m_i}{2\pi T_{i0}}\right)^{3/2}e^{-(p_\parallel^2/2m_i+\mu B)/T_{i0}},
\end{equation}
but it quickly evolves to a neoclassical equilibrium distribution function with steep plasma gradient.
The code uses two marker particle weights: time-invariant and time-varying. The time-invariant weights are used for initial marker particle loading in 5D phase space.  At $t>0$, the time-varying ion markers' weights evolve such that $F_i$ advances in time according to
\begin{equation}
\label{XGC_vlasov}
d_tF_i\doteq\pd_t F_i+\dot{\bd{R}}\cdot\nabla F_i+\dot{p}_\parallel \pd_{p_\parallel}F_i=C_i+S_i.
\end{equation}
Here, $C_i$ is a nonlinear Fokker--Planck--Landau collision operator \cite{Yoon14,Hager16}  and $S_i$ represents sources  (such as heating, torque input, and neutral ionization and charge exchange) and sinks (such as radiative heat loss).  In our simulations, only the effects from $C_i$ are retained, while $S_i$ is assumed to be zero. The wall perfectly absorbs ions. This is done by flipping the sign of $p_\parallel$ for the ion markers that hit the wall and changing their time-varying weights such that the outgoing $F_i=0$ \cite{Ku18}. 

Let us define the velocity-space integration as
\begin{equation}
\int d\mc{W}\doteq(2\pi/m_i^2)\int d\mu\,dp_\parallel B_\parallel^*.
\end{equation} 
The ion-density perturbation is then $\delta n_i\doteq\int d\mc{W} F_i-n_{i0}$,
and  $\Phi$ can be calculated from the gyrokinetic Poisson equation (with $Z_i=1$):
\begin{equation}
\label{XGC_poisson}
\nabla_\perp\cdot\left(\frac{n_{i0}m_i}{eB^2}\nabla_\perp\Phi\right)=-(\hat{J}_0\delta n_i-\delta n_e),
\end{equation}
where $\nabla_\perp$ denotes gradient perpendicular to $\bd{B}$.  For the electron-density perturbation $\delta n_e$, we adopt the adiabatic-electron model, which assumes 
\begin{equation}
\label{XGC_adia_elec}
\frac{\delta n_e}{n_{e0}}=\frac{e}{T_{e0}}\left(\Phi-\avg{\Phi}\right).
\end{equation}
Here, $n_{e0}(\psi)$ and $T_{e0}(\psi)$ are the equilibrium electron density and temperature, respectively. We have $n_{e0}=n_{i0}$ from charge neutrality, and  it is assumed that $T_{e0}=T_{i0}$. Also, $\avg{\dots}\doteq\int(\dots)\sqrt{g}\,d\theta/\int \sqrt{g}\,d\theta$ is the flux-surface average,
where $\sqrt{g}\doteq(\nabla\psi\times\nabla\varphi\cdot\nabla\theta)^{-1}$ is the Jacobian, and the integration over $\varphi$ is omitted due to the axisymmetry in XGCa. Note that for the SOL region, the flux-surface average  is performed along open (rather than closed) field lines between two wall-contacting points.

When solving the gyrokinetic Poisson equation~\eref{XGC_poisson}, the boundary condition is $\Phi=0$ at the wall, at the private-flux region below the X point, and  where $\psi_n>\psi_{\rm bdry}$. For the results presented in this article, we chose $\psi_{\rm bdry}=1.04$, but we also found that $E_r$ inside the LCFS is insensitive to the value of $\psi_{\rm bdry}$, as long as it is not too close to unity.  Since we used the adiabatic-electron model  \eref{XGC_adia_elec}, realistic electron dynamics and the sheath-boundary effects in the SOL are not included in our simulations. Kinetic electrons and sheath models should be included in the future in order to properly account for these effects \cite{Ku18}.

The electron radial current vanishes with the adiabatic-electron model \eref{XGC_adia_elec}. Then, the total ion radial current must also vanish due to the quasineutrality constraint, $J_r=J_{\rm pol} +J_{\rm gc}=0$. Therefore, to avoid confusion, we will refer to the ion radial gyrocenter flux $\Gamma_r\doteq J_{\rm gc}/Z_ie$ rather than the total ion radial current $J_r$ in the rest of the paper. Note that although $\Gamma_r$ can be nonzero, physically it is always balanced by the classical polarization flux $J_{\rm pol}/Z_ie$, which corresponds to the left-hand side of the gyrokinetic Poisson equation \eref{XGC_poisson}. This ensures that the total ion radial current is zero and the plasma is quasineutral.

Finally, we mention that our simulation setup is similar to a previous study of the plasma edge rotation using an earlier version of the code \cite{Chang08}, except that Ref.~\cite{Chang08} had kinetic electrons. 
As will be shown in \sref{edge_rotation}, results on the edge rotation from our simulations are qualitatively similar to those reported in Ref.~\cite{Chang08}, except that Ref.~\cite{Chang08} observed a radial electron current in the quasisteady state, while the adiabatic-electron model used here will set the radial electron current identically to zero. Since the edge rotation is not central to our study, we will not make quantitative comparisons.

\subsection{Ion-orbit-flux diagnostic implementation}
\label{formulation}
The ion-orbit-flux formulation uses the coordinates  $(\mu,\mc{P}_\varphi,\bar{H})$  to label ion gyrocenter orbits that cross the LCFS, where $\mc{P}_\varphi\doteq Z_ie\psi+p_\parallel\hat{\bd{b}}\cdot R^2\nabla\varphi$ is the canonical toroidal angular momentum and $\bar{H}$ is a chosen ``orbit Hamiltonian'' \cite{Stoltzfus20,Stoltzfus21}. The formulation is exact with any time-dependent $\bar{H}$,  provided that $\bar{H}$ is axisymmetric. For XGCa, one straightforward choice is $\bar{H}=H$,  but we will also make alternate choices to illustrate the formalism (\sref{core_diag}). The corresponding  orbit characteristics are given by
\begin{eqnarray}
\label{test}
\label{Rbardot}
B_\parallel^*\bar{\dot{\bd{R}}}=(Z_ie)^{-1}\hat{\bd{b}}\times\nabla\bar{H}+\bar{v}_\parallel \bd{B}^*,\\
\label{Pbardot}
B_\parallel^* \bar{\dot{p}}_\parallel=-\bd{B}^*\cdot\nabla \bar{H},
\end{eqnarray}
with $\bar{v}_\parallel\doteq\pd_{p_\parallel}\bar{H}$.  One can also define the ``remainder'' Hamiltonian $\tilde{H}\doteq H-\bar{H}$ and the corresponding quantities
\begin{eqnarray}
B_\parallel^*\tilde{\dot{\bd{R}}}=(Z_ie)^{-1}\hat{\bd{b}}\times\nabla\tilde{H}+\tilde{v}_\parallel \bd{B}^*,\\
B_\parallel^* \tilde{\dot{p}}_\parallel=-\bd{B}^*\cdot\nabla \tilde{H},
\end{eqnarray}
with $\tilde{v}_\parallel\doteq\pd_{p_\parallel}\tilde{H}$. (Note that $\tilde{v}_\parallel$ is always zero within this paper.) Let us define an orbit derivative at fixed time,
$\bar{d}_o\doteq\bar{\dot{\bd{R}}}\cdot\nabla+\bar{\dot{p}}_\parallel\pd_{p_\parallel}$ \cite{Stoltzfus21};
then, \eref{XGC_vlasov} becomes
\begin{equation}
\label{formulation_vlasov}
\bar{d}_o F_i=-\tilde{\dot{\bd{R}}}\cdot\nabla F_i-\tilde{\dot{p}}_\parallel\pd_{p_\parallel}F_i+C_i+S_i-\pd_t F_i.
\end{equation}
Note that  \eref{formulation_vlasov} is exactly equivalent to \eref{XGC_vlasov}, and we do not require $\tilde{H}$ to be smaller than $\bar{H}$. The  orbits $(\bar{\bd{R}}(\tau),\bar{p}_\parallel(\tau))$, parameterized by $\mu$, are obtained by integrating  \eref{Rbardot} and \eref{Pbardot} over a timelike variable $\tau$ at fixed true time $t$. It is straightforward to show that  $\bar{H}$ and $\mc{P}_\varphi$ are constant along the orbits \cite{Stoltzfus21}, namely,
\begin{equation}
\bar{d}_o\bar{H}=\bar{d}_o\mc{P}_\varphi=0.
\end{equation}
Therefore, the  orbits can be labeled by $(\mu,\mc{P}_\varphi,\bar{H})$, and they must form closed loops on the 2-dimensional plane $(R,z)$ \cite{Stoltzfus21}, unless they intersect the wall.  This means that any orbit that leaves the LCFS must have also entered it earlier. 

Using a coordinate transformation, the radial ion gyrocenter orbit flux through the LCFS can be expressed as \cite{Stoltzfus20}
\begin{equation}
\label{formulation_flux}
\eqalign{
\Gamma_r=\int d\bd{S}\cdot\int d\mc{W} F_i\bar{\dot{\bd{R}}}=\frac{2\pi}{Z_iem_i^2}\times
\\
\int_0^{\infty} d\mu\int_{-\infty}^{\infty} d\mc{P}_\varphi\sum_k\int_{\bar{H}^{\min}_k}^{\bar{H}^{\max}_k} d\bar{H}\oint d\varphi(F_i^{\rm out}-F_i^{\rm in}).
}
\end{equation}
Here, $d\bd{S}\doteq\sqrt{g}\,d\theta\,d\varphi\nabla\psi$ is the surface element.  The range of $\bar{H}$ is determined by varying $\theta$ from $0$ to $2\pi$ at fixed $\mu$ and $\mc{P}_\varphi$, and $k$ labels the local minimum and maximum of $\bar{H}(\theta)$. For a given orbit, $F_i^{\rm in}$ and $F_i^{\rm out}$ are evaluated at the incoming and the outgoing points where the orbit crosses the LCFS. 
The difference between $F_i^{\rm in}$ and $F_i^{\rm out}$ can be calculated by integrating the orbit derivative along the orbits, \ie
\begin{equation}
\label{formulation_flux2}
F^{\rm out}_i-F^{\rm in}_i=\int_0^{\tau_{\rm orb}}d\tau\,\bar{d}_o F_i.
\end{equation} 
Then, we have our ion-orbit-flux formulation ready for numerical evaluations:
\begin{equation}
\label{formulation_final}
\eqalign{
\Gamma_r=\frac{2\pi}{Z_iem_i^2}
\int_0^{\infty} d\mu\int_{-\infty}^{\infty} d\mc{P}_\varphi
\sum_k\int_{\bar{H}^{\min}_k}^{\bar{H}^{\max}_k} d\bar{H}
\oint d\varphi\times
\\
\int_0^{\tau_{\rm orb}}d\tau(-\tilde{\dot{\bd{R}}}\cdot\nabla F_i-\tilde{\dot{p}}_\parallel\pd_{p_\parallel}F_i+C_i+S_i-\pd_t F_i)
\\
\doteq\Gamma_{\rm rem}+\Gamma_{\rm col}+\Gamma_{\rm s}+\Gamma_t.
}
\end{equation}
Here, the remainder flux $\Gamma_{\rm rem}$ is from  $-\tilde{\dot{\bd{R}}}\cdot\nabla F_i-\tilde{\dot{p}}_\parallel\pd_{p_\parallel}F_i$, the collisional flux $\Gamma_{\rm col}$ is from  $C_i$, the source flux $\Gamma_{\rm s}$ is from  $S_i$, and $\Gamma_t$ is from $-\pd_t F_i$. The term $\Gamma_{\rm rem}$ arises whenever the remainder Hamiltonian $\tilde{H}$ is nonzero. For example, this would include effects from turbulence, when $\tilde{H}$ is non-axisymmetric. But $\Gamma_{\rm rem}$ can also include other effects. For example, in \sref{core_diag2}, $\tilde{H}$ varies poloidally but is still axisymmetric. Then, the corresponding $\Gamma_{\rm rem}$ describes effects from neoclassical processes rather than from turbulence. Meanwhile, note that $\Gamma_t$ is not a transport or source term \emph{per se}; rather, a nonzero $\Gamma_t$ is the result of imbalance between transport and sources along the orbits that give rise to Eulerian time change in $F_i$. 
 
To study steady-state transport and sources onto the loss orbits, we also define a loss-region function $L(\mu,\mc{P}_\varphi,\bar{H})$, such that $L=1$ for the loss orbits, which intersect the wall, and $L=0$ for the confined orbits, which do not intersect the wall. Then, by inserting $L$ into the integrand of \eref{formulation_final}, we get the loss-orbit contribution to the flux; namely, the remainder flux can be decomposed into the loss-orbit contribution and the confined-orbit contribution:
\begin{equation}
\label{formulation_loss}
\Gamma_{\rm rem}=\Gamma_{\rm rem}^{\rm loss}+\Gamma_{\rm rem}^{\rm conf},
\end{equation}
and similarly for $\Gamma_{\rm col}$, $\Gamma_{\rm s}$, and $\Gamma_t$. Note that when evaluating $F_i^{\rm out}-F_i^{\rm in}$ from \eref{formulation_flux2} for confined orbits, the integration path can be either the confined-region part of the orbit inside the LCFS, or the SOL part outside the LCFS. (Mathematically they yield the same results.) However, for loss orbits, the integration path can only be the confined-region part, as the SOL part will intersect the wall. It may therefore appear that the formalism misses effects from transport and sources along the SOL part of the loss orbits, e.g., scattering of loss-orbit particles from the SOL back into the confined region. However, their contribution is still entirely retained via $F_i^{\rm in}$. Specifically, in the SOL, transport and sources from the confined orbits to the loss orbits will result in nonzero $F_i^{\rm in}$ of the loss orbits, which then contributes to $\Gamma^{\rm loss}$. Similarly, transport and sources from the loss orbits to the confined orbits will change $F_i^{\rm in}$ of the confined orbits, which then contributes to $\Gamma^{\rm conf}$.

Finally, we note that the above formulations apply to any closed flux surface, not just the LCFS. For example, in \sref{core} below, the formulations are applied to a core flux surface $\psi_n=0.4$.

The orbit-flux formulation \eref{formulation_final} has been numerically implemented  in XGCa. A separate ion-orbit code has been developed \cite{ion-orbit}, which describes the orbits $(\bar{\bd{R}}(\tau),\bar{p}_\parallel(\tau))$ labeled by $(\mu,\mc{P}_\varphi,\bar{H})$. Then, $F_i^{\rm out}-F_i^{\rm in}$ is calculated in XGCa according to \eref{formulation_flux2}. Finally, the orbit flux is obtained from \eref{formulation_flux}, where $\oint d\varphi$ is replaced with $2\pi$ for XGCa. The ion-orbit code also determines the loss-orbit region with the function $L(\mu,\mc{P}_\varphi,\bar{H})$, so the loss-orbit contribution to the flux can be determined as in \eref{formulation_loss}. To avoid repetitive XGCa simulations, we also made an alternative implementation where XGCa outputs $F_i$, $C_i$, and $S_i$, then a separate code calculates the orbit flux \cite{orbit-flux}.  

\section{Collisional relaxation of core $E_r$}
\label{core}
In this section, we demonstrate the validity of our ion-orbit-flux diagnostic by studying the neoclassical $E_r$ at a core flux surface $\psi_n=0.4$. Results in this section are from the forward-$\nabla B$ configuration with $B_\varphi<0$. Results for the backward-$\nabla B$ configuration are similar. 
\subsection{Plasma properties in quasisteady state}
\begin{figure}
\includegraphics[width=1\columnwidth]{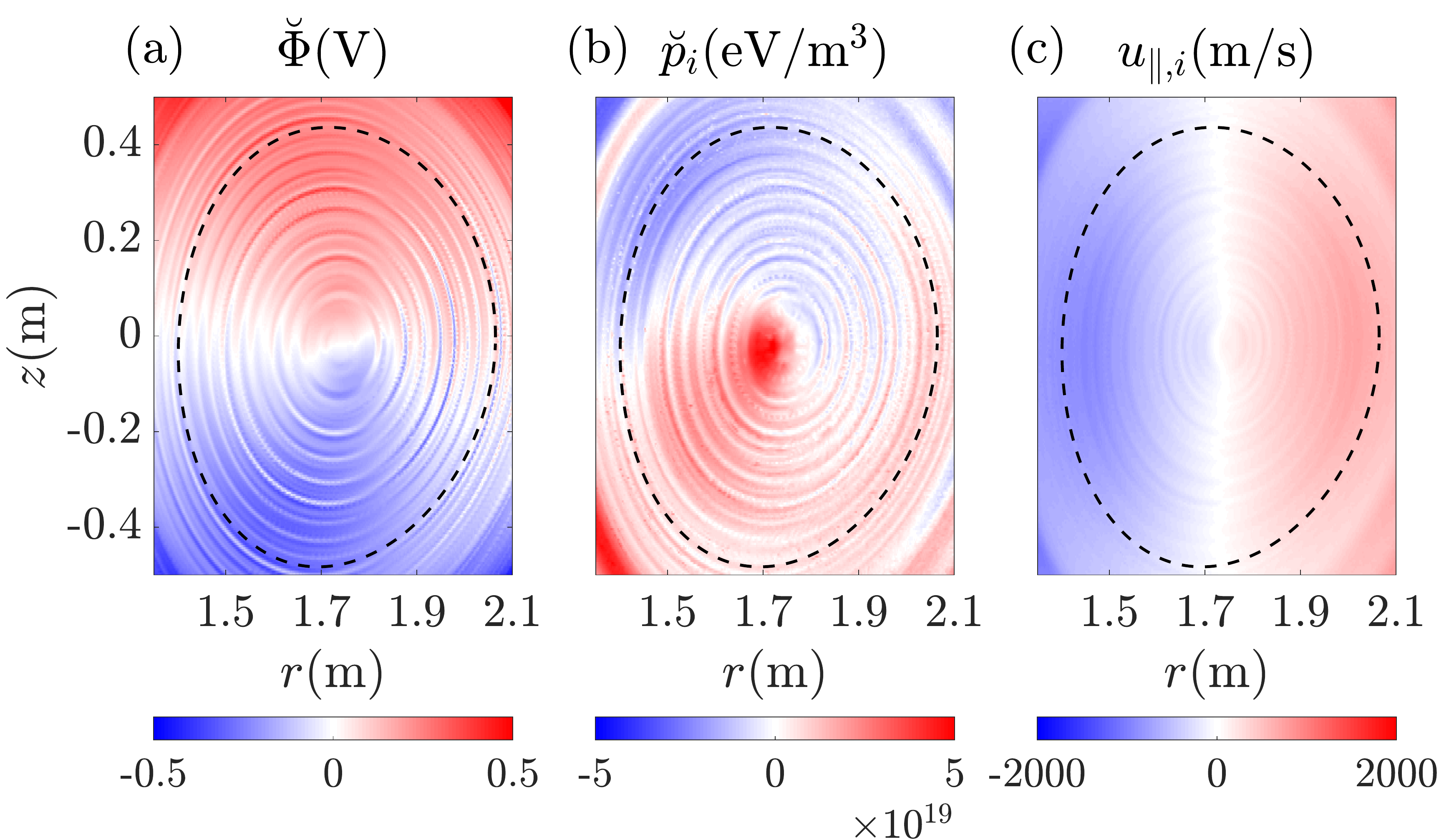}\caption{The nonzonal electrostatic potential perturbation $\breve{\Phi}$ (a), the nonzonal ion temperature perturbation $\breve{p}_i$ (b), and the ion fluid parallel velocity $u_{\parallel,i}$ (c). The data is averaged over 1.2ms$<t<$1.6ms to reduce numerical noise. The black dashed curve shows the $\psi_n=0.4$ flux surface.}\label{core2d}
\end{figure}

The initially Maxwellian distribution function does not correspond to a neoclassical equilibrium, so geodesic-acoustic-mode (GAM) oscillations are excited. However, the GAM oscillations are quickly damped by Coulomb collisions and ion Landau damping, after which the system reaches a quasisteady state.  \Fref{core2d} shows the nonzonal potential perturbation $\breve{\Phi}\doteq\Phi-\avg{\Phi}$, the nonzonal ion-pressure perturbation $\breve{p}_i\doteq p_i-\avg{p_i}$ where $p_i\doteq\int d\mc{W}(\mu B+p_\parallel^2/2m_i)F_i$,  and the ion fluid parallel velocity $u_{\parallel,i}\doteq\int d\mc{W}v_\parallel F_i/\int d\mc{W}F_i$ in the quasisteady state.  (We use the breve notation here to denote nonzonal quantities, because the more standard tilde notation has already been used in \sref{formulation} for other purposes.) Since we focus on the core flux surface $\psi_n=0.4$ in this section, we only show these field quantities around this surface. It is seen that both $\breve{\Phi}$ and $\breve{p}_i$ have up-down dipolar structures but they are out of phase, so the parallel electric field is against the ion parallel pressure gradient. Note that the nonzonal ion-gyrocenter density perturbation $\breve{n}_i$ is in phase with $\breve{\Phi}$ because of the adiabatic-electron model \eref{XGC_adia_elec}, assuming $\breve{n}_i\approx\breve{n}_e$ in the core. Therefore, $\breve{n}_i$ and $\breve{p}_i$ are also out of phase, due to a nonzonal ion-temperature perturbation out of phase with $\breve{n}_i$.  Meanwhile, $u_{\parallel,i}$ has a left-right dipolar structure, so the plasma flows from the bottom ($z<0$) to the top ($z>0$).  Such a $u_{\parallel,i}$ would cause a density buildup at the top, but the perpendicular flow makes the total plasma flow divergence-free.

\begin{figure}
\includegraphics[width=1\columnwidth]{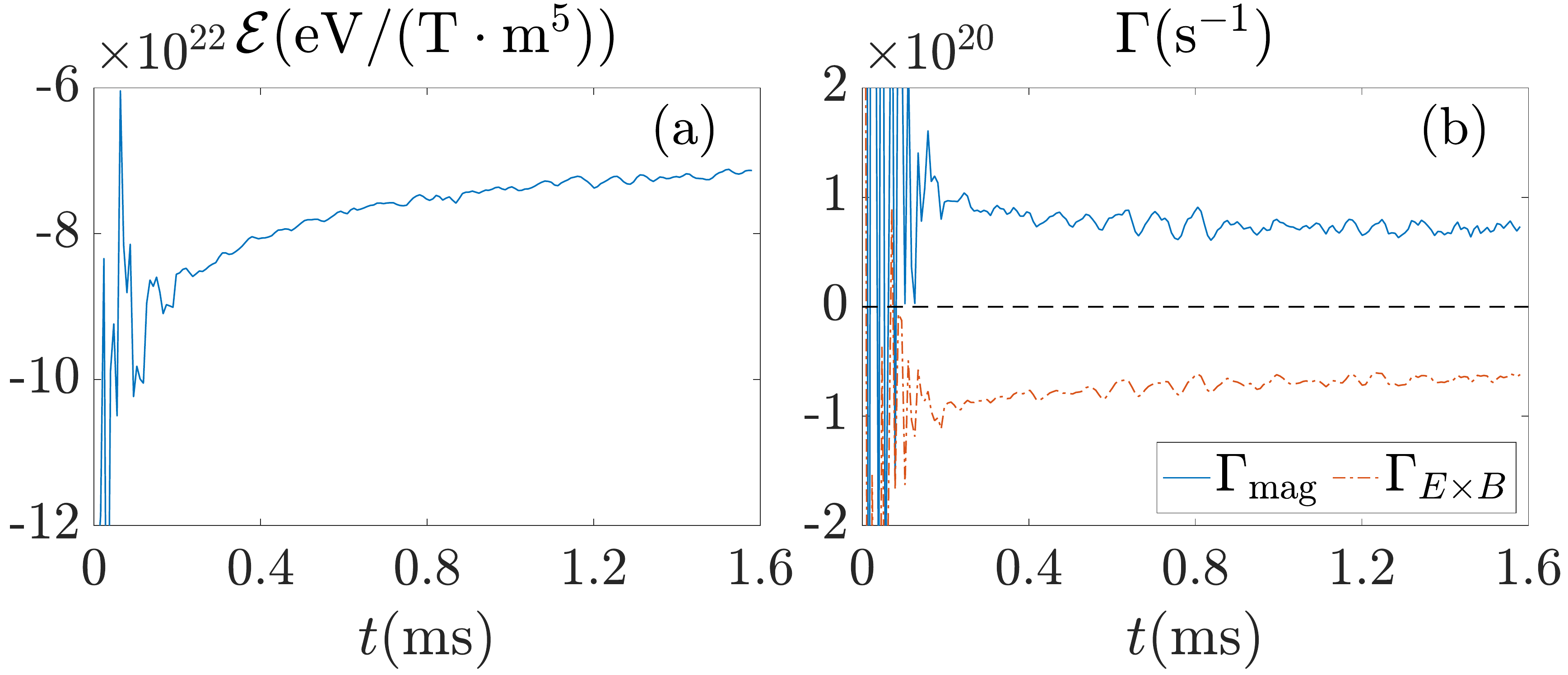}\caption{(a) The quantity $\mc{E}$  \eref{core_E} versus $t$ at the $\psi_n=0.4$ flux surface. (b) The ion radial gyrocenter  magnetic flux  (blue solid curve) and $\ExB$ flux  (red dot-dashed curve) across the same flux surface. To reduce numerical fluctuations, the data has been smoothed by averaging over a time window of $\Delta t=0.04$ms.}\label{coreEr}
\end{figure}

Since $E_r=-\nabla\Phi\cdot\nabla\psi/|\nabla \psi|$ is not a flux function, we look at the following quantity
\begin{equation}
\label{core_E}
\mc{E}(\psi,t)\doteq-en_{i0}\pd_\psi\avg{\Phi},
\end{equation}
which can be considered as the radial electric force on ions from the zonal $E_r$. In the following, we use zonal $E_r$ and $\mc{E}$ interchangeably. \Fref{coreEr}(a) shows the time evolution of $\mc{E}$ at the $\psi_n=0.4$ surface.  After the initial GAM oscillations, $\mc{E}$ goes through a slower collisional damping process until it relaxes to its neoclassical value. The gyrokinetic Poisson equation \eref{XGC_poisson} relates the zonal $E_r$ with the volume-integrated $\hat{J}_0\delta n_i$. Therefore, from the continuity relation, the ion radial gyrocenter flux must also vanish as $\mc{E}$ approaches a constant value in the present adiabatic electron model applied to the core plasma. 

Let us separate the radial gyrocenter flux into the magnetic-drift part and the $\ExB$-drift part:
\begin{equation}
\label{core_flux}
\int d\bd{S}\cdot\int d\mc{W}F_i\dot{\bd{R}}=\Gamma_{\rm mag}+\Gamma_{\exb},
\end{equation}
where  $\Gamma_{\rm mag}\doteq\int d\bd{S}\cdot\int d\mc{W} B_\parallel^{*-1} F_i(\hat{\bd{b}}\times\mu\nabla  B/e+v_\parallel\bd{B}^*)$ and  $\Gamma_{\exb}\doteq\int d\bd{S}\cdot\int d\mc{W} B_\parallel^{*-1} F_i(\hat{\bd{b}}\times\nabla \hat{J}_0\Phi)$. \Fref{coreEr}(b) shows that $\Gamma_{\rm mag}>0$ and $\Gamma_{\exb}<0$ in the quasisteady state, such that the total flux is zero.  These results are consistent with the dipolar structures of $\breve{p}_i$ and $\breve{\Phi}$ shown in \fref{core2d}. Namely, since $\breve{p}_i>0$ for $z<0$  and the magnetic drift points in the negative-$z$ direction, the net magnetic flux is radially outward, $\Gamma_{\rm mag}>0$. Meanwhile,  the $\ExB$ drift points in the negative-$R$ direction, and is stronger at the outboard side ($\theta\approx 0$) than the inboard side ($\theta\approx\pi$); hence, the net $\ExB$ flux is radially inward, $\Gamma_\exb<0$.

\subsection{Ion-orbit-flux diagnostic results}
\label{core_diag}
Below, we show the orbit-flux diagnostic results using two different orbit Hamiltonians: $\bar{H}=\bar{H}_1\doteq p_\parallel^2/2m_i+\mu B+\hat{J}_0e\Phi$ (\sref{core_diag1}) and $\bar{H}=\bar{H}_2\doteq p_\parallel^2/2m_i+\mu B+\hat{J}_0e\avg{\Phi}$ (\sref{core_diag2}). 
With $\bar{H}=\bar{H}_1$, we will show that the results of $\Gamma_{\rm col}$ and $\Gamma_t$ are consistent with the collisional relaxation of $E_r$. With $\bar{H}=\bar{H}_2$, the orbit flux also contains a nonzero $\Gamma_{\rm rem}$  due to the nonzero $\tilde{H}=H-\bar{H}$. The physical meaning of $\Gamma_{\rm rem}$ will be briefly discussed in this section. The source flux  $\Gamma_{\rm s}$ will  not be considered, since $S_i$ is set to zero.
\subsubsection{Results with $\bar{H}=\bar{H}_1$.}
\label{core_diag1}
Since $\bar{H}_1= p_\parallel^2/2m_i+\mu B+\hat{J}_0e\Phi$ equals the true Hamiltonian $H$, the orbit characteristics $\bar{\dot{\bd{R}}}=\bar{\dot{\bd{R}}}_1$ (governed by \eref{Rbardot}) are the same as the true characteristics $\dot{\bd{R}}$ (governed by \eref{XGC_Rdot}). Then, comparing  \eref{formulation_final} and \eref{core_flux}, the orbit flux is
\begin{equation}
\label{core_diag_flux1}
\int d\bd{S}\cdot\int d\mc{W}F_i\bar{\dot{\bd{R}}}_1=\Gamma_{\rm mag}+\Gamma_{\exb}=\Gamma_{\rm col}^1+\Gamma_{t}^1,
\end{equation}
where we put a superscript ``1'' on $\Gamma_{{\rm col},t}$ (meaning $\Gamma_{\rm col}$ or $\Gamma_t$) to indicate that they are evaluated along $\bd{R}=\bar{\bd{R}}_1$.
Since $\mc{E}<0$ and $|\mc{E}|$ decreases during the collisional relaxation (\fref{coreEr}(a)), the net radial flux is negative, so we expect $\Gamma_{\rm col}^1+\Gamma_{t}^1$ to be negative. This expectation is confirmed by the results shown in  \fref{coreFlux}(a).  The collisional flux $\Gamma_{\rm col}^1$ is negative, corresponding to a radially inward current induced by viscosity. The term $\Gamma_{t}^1$ is positive, but the net radial flux $\Gamma_{\rm col}^1+\Gamma_{t}^1$ is negative and damps to zero, consistent with the collisional relaxation of $\mc{E}$.
\subsubsection{Results with $\bar{H}=\bar{H}_2$.}
\label{core_diag2}
To further explore the formulation, we can alternatively choose $\bar{H}=\bar{H}_2\doteq p_\parallel^2/2m_i+\mu B+\hat{J}_0e\avg{\Phi}$, which only includes the zonal part of $\Phi$. Since $\bar{H}_2\neq H$, the orbit characteristics $\bar{\dot{\bd{R}}}=\bar{\dot{\bd{R}}}_2$ (governed by \eref{Rbardot}) are different from the true characteristics $\dot{\bd{R}}$ (governed by \eref{XGC_Rdot}). Specifically, the magnetic drifts in $\bar{\dot{\bd{R}}}_2$ and  $\dot{\bd{R}}$ are the same, but the $\ExB$ drift in $\bar{\dot{\bd{R}}}_2$ is always tangent to flux surfaces (assuming $\hat{J}_0\approx 1$ in the core). Therefore, the $\ExB$-drift contribution to the orbit flux vanishes, leaving only the magnetic-drift contribution:
\begin{equation}
\label{core_diag_flux2}
\int d\bd{S}\cdot\int d\mc{W}F_i\bar{\dot{\bd{R}}}_2=\Gamma_{\rm mag}=\Gamma_{\rm rem}+\Gamma_{\rm col}^2+\Gamma_{t}^2.
\end{equation}
Here,  $\Gamma_{\rm rem}$ comes from the remainder Hamiltonian $\tilde{H}=H-\bar{H}_2\approx e\breve{\Phi}$. For emphasis, the nonzonal potential $\breve{\Phi}$  varies along the poloidal direction, but is still axisymmetric.  We  put a superscript ``2'' on $\Gamma_{{\rm col},t}$ to indicate that they are evaluated along $\bd{R}=\bar{\bd{R}}_2$, which is different from $\bar{\bd{R}}_1$ since $\bar{H}_2\neq\bar{H}_1$.

\begin{figure}
\centering
\includegraphics[width=1\columnwidth]{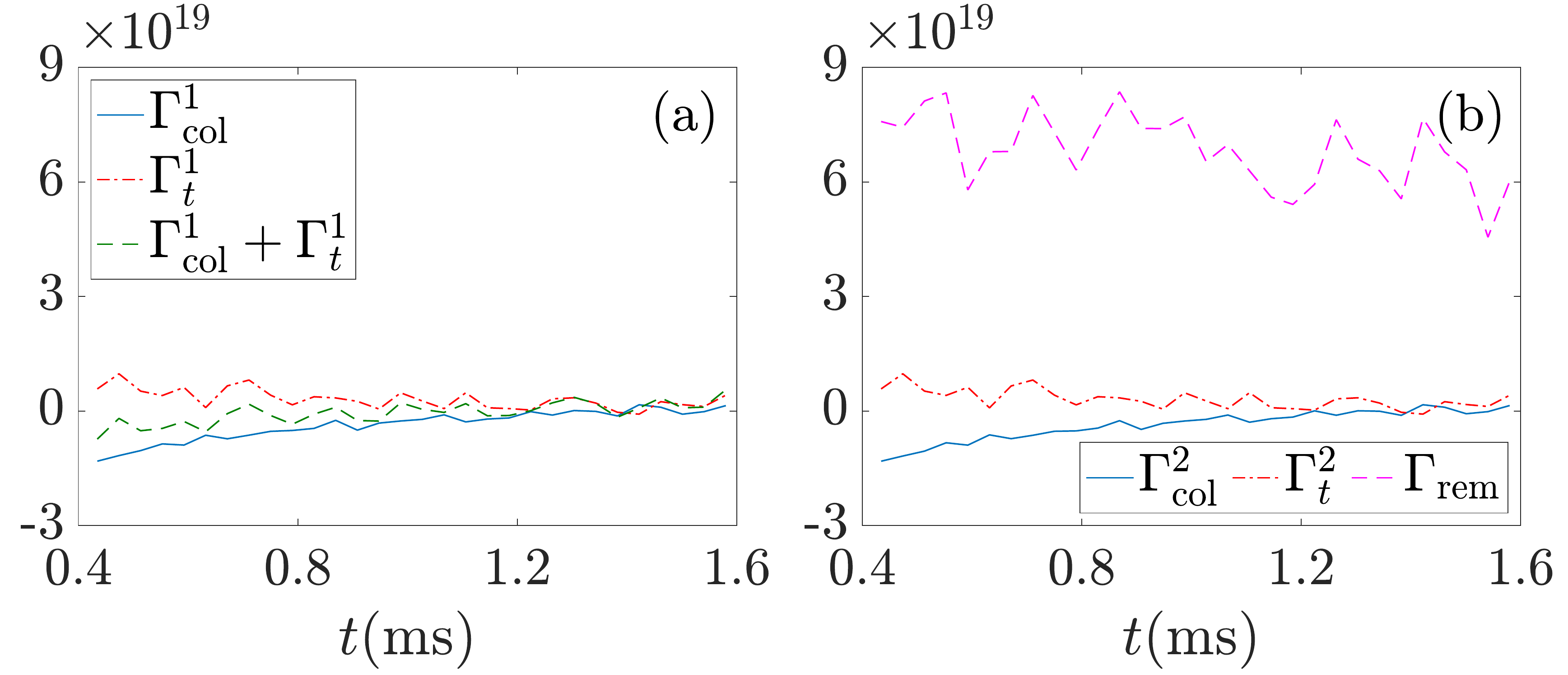}\caption{The orbit fluxes (in units ${\rm s}^{-1}$) at the $\psi_n=0.4$  surface, with two different choices of the orbit Hamiltonian: (a) $\bar{H}=\bar{H}_1$ (\sref{core_diag1}), and (b) $\bar{H}=\bar{H}_2$ (\sref{core_diag2}). In (a), shown are $\Gamma_{\rm col}^1$ (blue solid curve), $\Gamma_t^1$ (red dot-dashed curve), and  the sum of the two terms (green dashed curve). In (b), shown are $\bar{\Gamma}_{\rm col}^2$ (blue solid curve), $\bar{\Gamma}_t^2$ (red dot-dashed curve), and  $\Gamma_{\rm rem}$ (magenta dashed curve). To reduce numerical fluctuations, $\Gamma_t$ has been smoothed by averaging over a time window of $\Delta t=0.08$ms.}\label{coreFlux}
\end{figure}

\Fref{coreFlux}(b) shows the orbit-flux diagnostic results with $\bar{H}=\bar{H}_2$. Because $e\breve{\Phi}\ll T_{i0}$ in the core (figures \ref{equilibrium}(d) and \ref{core2d}(a)), the difference between $\bar{H}_1$ and $\bar{H}_2$ is tiny, and $\bar{\bd{R}}_1$ and $\bar{\bd{R}}_2$ are almost identical. Consequently, there is no visible difference between $\Gamma_{{\rm col},t}^2$ in \fref{coreFlux}(b) and $\Gamma_{{\rm col},t}^1$ in \fref{coreFlux}(a).  With $\Gamma_{{\rm col},t}^1\approx \Gamma_{{\rm col},t}^2$, and by comparing \eref{core_diag_flux1} and \eref{core_diag_flux2}, we expect $\Gamma_{\rm rem}\approx -\Gamma_{\exb}$. This expectation is also confirmed by the results shown in \fref{coreFlux}(b), namely, $\Gamma_{\rm rem}\approx 7\times 10^{19}{\rm s}^{-1}$ is approximately the same as $-\Gamma_{\exb}$ in \fref{coreEr}(b).

In this case, most of $\Gamma_{\rm rem}$ comes from the  acceleration term $-\tilde{\dot{p}}_\parallel\pd_{p_\parallel}F_i$, while the contribution of the advection term $-\tilde{\dot{\bd{R}}}\cdot\nabla F_i$ is small and fluctuates around zero. This finding is consistent with the ordering estimate that $\tilde{\dot{\bd{R}}}\cdot\nabla F_i\sim E_\theta F_i/BL$ and  $\tilde{\dot{p}}_\parallel\pd_{p_\parallel}F_i\sim eE_\theta B_\theta F_i/m_i v_t B$, so the ratio between the two terms is $\rho_{i\theta}/L$, which is very small in the core. Here,  $L$ is an equilibrium length scale, $v_t$ is ions' thermal speed, and $\rho_{i\theta}\doteq m_iv_t/eB_\theta$ is the poloidal gyroradius.   For emphasis, this ordering is specific to our axisymmetric system in the core, and would not hold if $\breve{\Phi}$ contained turbulent fluctuations.

The fact that $\Gamma_{\rm rem}\approx -\Gamma_{\exb}$ is nontrivial. Both fluxes stem from the nonzero $\breve{\Phi}$, but $\Gamma_{\exb}$ is evaluated at the surface, while $\Gamma_{\rm rem}$ is evaluated along the orbits that cross the surface. Further, since $\Gamma_{\rm mag} +\Gamma_\exb\approx 0$ in steady state, we have $\Gamma_{\rm mag}\approx \Gamma_{\rm rem}$. Recalling  \eref{formulation_flux2} and \eref{formulation_vlasov}, this reveals that the steady-state positive $\Gamma_{\rm mag}$ is supported by the acceleration from  the parallel electric field $E_\parallel=-\hat{\bd{b}}\cdot\nabla\breve{\Phi}$.  Specifically, since the magnetic drift points in the negative-$z$ direction, orbits' incoming points are at the top of the flux surface and the outgoing points are at the bottom. Co-current ions ($v_\parallel>0$) move counterclockwise to the inboard side where $E_\parallel>0$, so the corresponding parallel acceleration term $-\tilde{\dot{p}}_\parallel\pd_{p_\parallel}F_i$ is positive, assuming $\pd_{p_\parallel}F_i<0$ at $v_\parallel>0$. Similarly, counter-current ions ($v_\parallel<0$) move to the outboard side where $E_\parallel<0$, so $-\tilde{\dot{p}}_\parallel\pd_{p_\parallel}F_i$ is also positive for them. Therefore, $\breve{p}_i>0$ at the bottom (\fref{core2d}(b)) and the net magnetic-drift flux is positive, $\Gamma_{\rm mag}>0$.
\section{Effects of ion orbit loss on edge $E_r$}
\label{edge}
\begin{figure}
\includegraphics[width=1\columnwidth]{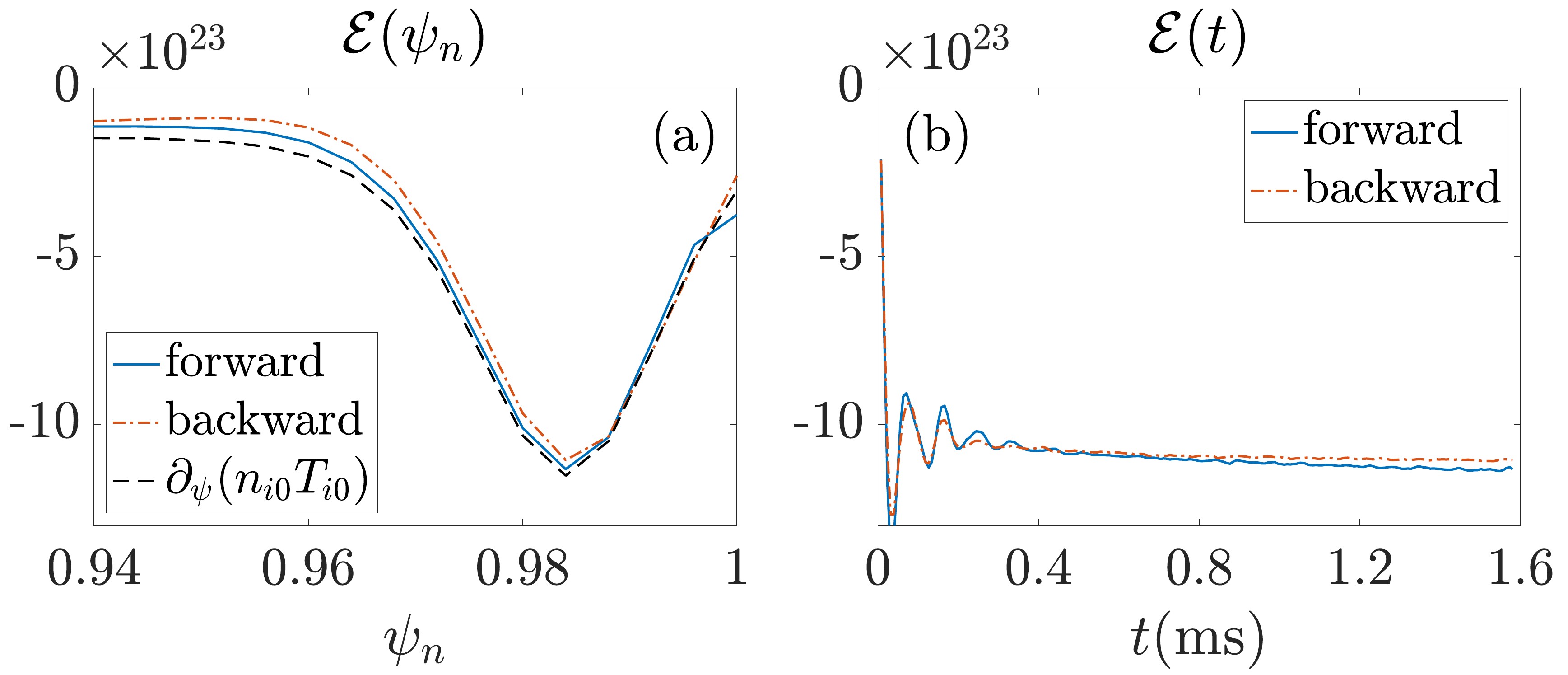}\caption{(a)  $\mc{E}$ versus $\psi_n$ at $t=1.6{\rm ms}$ from the forward-$\nabla B$ (blue solid curve) and the backward-$\nabla B$ configuration (red dot-dashed curve). Also shown is the radial ($\psi$) gradient of the equilibrium ion pressure (black dashed curve). (b) $\mc{E}$ at $\psi_n=0.984$ versus $t$ from the two configurations.}\label{edgeEr}
\end{figure}

In this section, the orbit-flux diagnostic is used to study the effects of ion orbit loss on the edge $E_r$. Results from the forward- and backward-$\nabla B$ configurations are compared. It is found that the radial electric force on ions from $E_r$ approximately balances the ion radial pressure gradient for both configurations, indicating that the radial $\bd{v}\times \bd{B}$ force on ions plays only a minor role. For the orbit-flux diagnostic results, both $\Gamma_{\rm col}$ and $\Gamma_t$ are nonzero but they cancel each other. The loss-orbit contribution to the flux is estimated to be $\Gamma_r^{\rm loss}\approx 10^{19}{\rm s}^{-1}$, which is found to be able to drive $E_r$ away by a few percent from its standard neoclassical solution. The nonzero $\Gamma_t$ is related to a toroidal-rotation acceleration in the edge that persists even when $E_r$ is quasisteady.

\subsection{Ion-orbit-flux diagnostic results}
\Fref{edgeEr} shows $\mc{E}$ at the edge. A large $E_r$ well is developed inside the LCFS, similar to experimental observations. The resulting radial electric force on ions approximately balances the radial ion pressure gradient caused by the density and temperature pedestal shown in \fref{equilibrium}(c). For the forward-$\nabla B$ case, the value of $\mc{E}$ at the trough slowly grows more negative in quasisteady state (\fref{edgeEr}(b)). For the backward-$\nabla B$ case, $\mc{E}$ also shifts in the negative direction, but even more slowly. The increase of $|\mc{E}|$ for the forward-$\nabla B$ case indicates a small radially outward gyrocenter ion flux, which is estimated to be $\Gamma_r\approx 10^{19}{\rm s}^{-1}$; for the backward-$\nabla B$ case, $\Gamma_r$ is even smaller.

For the orbit-flux diagnostic, we choose $\bar{H}=H$ so the orbit fluxes consist of  $\Gamma_{\rm col}$ and $\Gamma_t$. The loss-orbit contribution to the fluxes is also calculated using the loss-region function $L$, which is numerically determined by the ion-orbit code \cite{ion-orbit}. The loss orbits make up about 30\% of all the orbits considered, that is $\int L\,d\mu\,d\mc{P}_\varphi\, d\bar{H}/\int d\mu \,d\mc{P}_\varphi\, d\bar{H}\approx 0.3$. This percentage is not significantly different between the forward- and backward-$\nabla B$ configurations, since the two cases have the same magnetic-field topology and similar levels of $E_r$. We have also verified that the shape of the loss region from our numerical $L$ (not shown) is consistent with some earlier analytic studies \cite{deGrassie09,Chankin93,Miyamoto96}. 

\begin{figure}
\includegraphics[width=1\columnwidth]{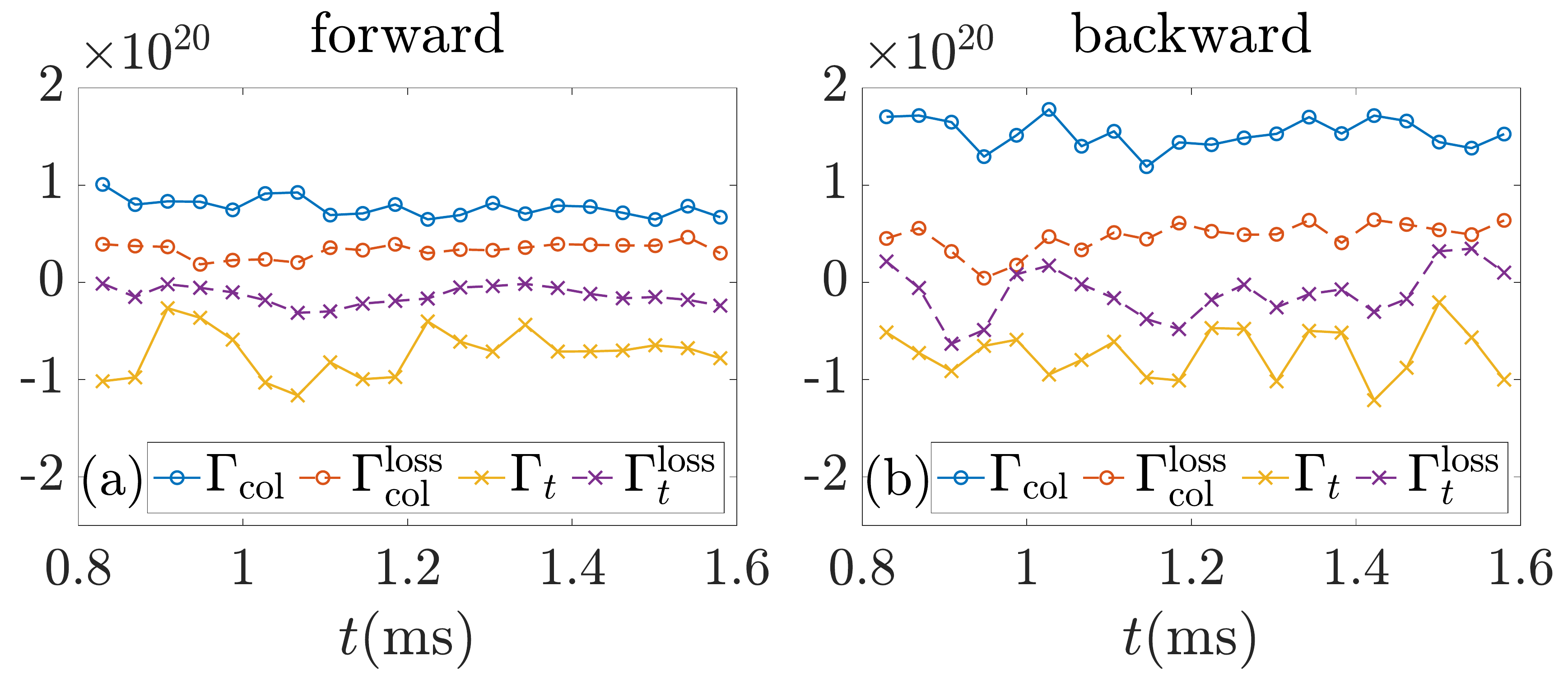}\caption{The orbit-flux diagnostic results at the LCFS.  Shown are the results of $\Gamma_{\rm col}$ (circles) and $\Gamma_{t}$ (crosses) for the forward-$\nabla B$  (a) and the backward-$\nabla B$ configurations (b). The solid curves are the total fluxes including all orbits, while the dashed curves are contributions from the loss orbits alone. To reduce numerical fluctuations, $\Gamma_t$ has been smoothed by averaging over a time window of $\Delta t=0.08$ms.}\label{edgeFlux}
\end{figure}

\Fref{edgeFlux} shows the orbit-flux diagnostic results at the LCFS ($\psi_n=1$). Unlike the core (\fref{coreFlux}), here at the edge $\Gamma_{\rm col}$ and $\Gamma_t$ do not vanish in quasisteady state, and their magnitudes are both large, $|\Gamma_{{\rm col},t}|\approx 10^{20}{\rm s}^{-1}$. But $\Gamma_{\rm col}>0$ and $\Gamma_t<0$, so the total orbit flux $\Gamma_{\rm col}+\Gamma_{t}$ is much smaller. Similarly, for the loss-orbit contribution,  $\Gamma_{\rm col}^{\rm loss}>0$ and $\Gamma_{t}^{\rm loss}<0$. Due to the cancellation between $\Gamma_{\rm col}^{\rm loss}$ and $\Gamma_{t}^{\rm loss}$, the net loss-orbit flux appears to be small, making its accurate evaluation difficult in the presence of numerical fluctuations. For this reason, we only provide an estimate of the loss-orbit flux, which is $\Gamma_r^{\rm loss}\approx 10^{19}{\rm s}^{-1}$. The magnitudes of $\Gamma_r^{\rm loss}$ are similar between the forward-$\nabla B$  and the backward-$\nabla B$ configurations. The small value of $\Gamma_r^{\rm loss}$ could be a result of the large $|E_r|$ providing good magnetoelectric confinement of ions \cite{Stix71}.

Note that $\Gamma_r^{\rm loss}\approx 10^{19}{\rm s}^{-1}$ is on the same level as the estimated net ion gyrocenter flux $\Gamma_r$ for the forward-$\nabla B$ configuration. Therefore,  $\Gamma_r^{\rm loss}$ may not be entirely balanced by the confined-orbit flux $\Gamma_r^{\rm conf}$ in quasisteady state for this case. For the backward-$\nabla B$ configuration, however, the estimated $\Gamma_r$ is much smaller than $\Gamma_r^{\rm loss}$; hence, the assumption that $\Gamma_r^{\rm conf}$ balances $\Gamma_r^{\rm loss}$ works better for this case.
\subsection{Effects of $\Gamma_r^{\rm loss}$ on $E_r$}
Let us integrate the gyrokinetic Poisson equation \eref{XGC_poisson} over the volume inside the LCFS. From Gauss's law, we have 
\begin{equation}
\label{edge_gauss}
\int \frac{n_{i0}m_i}{eB^2} E_r\,dS=\int \hat{J}_0\delta n_i\,dV,
\end{equation}
where $dS=|d\bd{S}|$ and $dV$ is the volume element. Since the right-hand side of \eref{edge_gauss}
is approximately the perturbation of number of gyrocenter ions inside the LCFS, we take the time derivative of \eref{edge_gauss} and use the continuity equation. This gives a relation between the ion radial gyrocenter flux and the radial electric field:
\begin{equation}
\label{edge_flux1}
\Gamma_r\approx - \frac{\pd}{\pd t}\int \frac{n_{i0}m_i}{eB^2} E_r\,dS.
\end{equation}
(The electron radial flux is zero due to the adiabatic-electron assumption.) The right hand side represents the ion radial polarization flux.
We can further separate $\Gamma_r$ into the confined-orbit contribution and the loss-orbit contribution, $\Gamma_r=\Gamma_r^{\rm conf}+\Gamma_r^{\rm loss}$. Without orbit loss, $\Gamma_r=\Gamma_r^{\rm conf}$, and $E_r$ would approach its standard neoclassical solution $E_r^{\rm neo(0)}$, which is determined by \cite{Hinton73,Hazeltine74,Hirshman78,Hirshman81}
\begin{equation}
\Gamma_r^{\rm conf}(E_r^{\rm neo(0)})=0.
\end{equation}
Suppose a nonzero $\Gamma_r^{\rm loss}$ leads to $E_r=E_r^{\rm neo(0)}+\Delta E_r$ in steady state.  We then have
\begin{equation}
\label{edge_flux2}
\Gamma_r^{\rm conf}(E_r^{\rm neo(0)}+\Delta E_r)=-\Gamma_r^{\rm loss}.
\end{equation}
For simplicity, let us assume that $\Gamma_r^{\rm conf}$ damps $E_r$ towards $E_r^{\rm neo(0)}$ with a  neoclassical collisional poloidal-rotation damping rate $\nu_p$. Then, from \eref{edge_flux1} and \eref{edge_flux2}, $\Delta E_r$ is estimated from the following relation:
\begin{equation}
\nu_p\int \frac{n_{i0}m_i}{eB^2}\Delta E_r\,d{S}\approx -\Gamma^{\rm loss}_r.
\end{equation}
The ${n_{i0}m_i}/{eB^2}$ factor in the integrand on the left hand side follows from the dielectric response of a homogeneous magnetized plasma, while $\nu_p$ is due to collisional poloidal-rotation damping by magnetic inhomogeneity in a tokamak plasma \cite{Hirshman78}. 

\begin{figure}
\includegraphics[width=1\columnwidth]{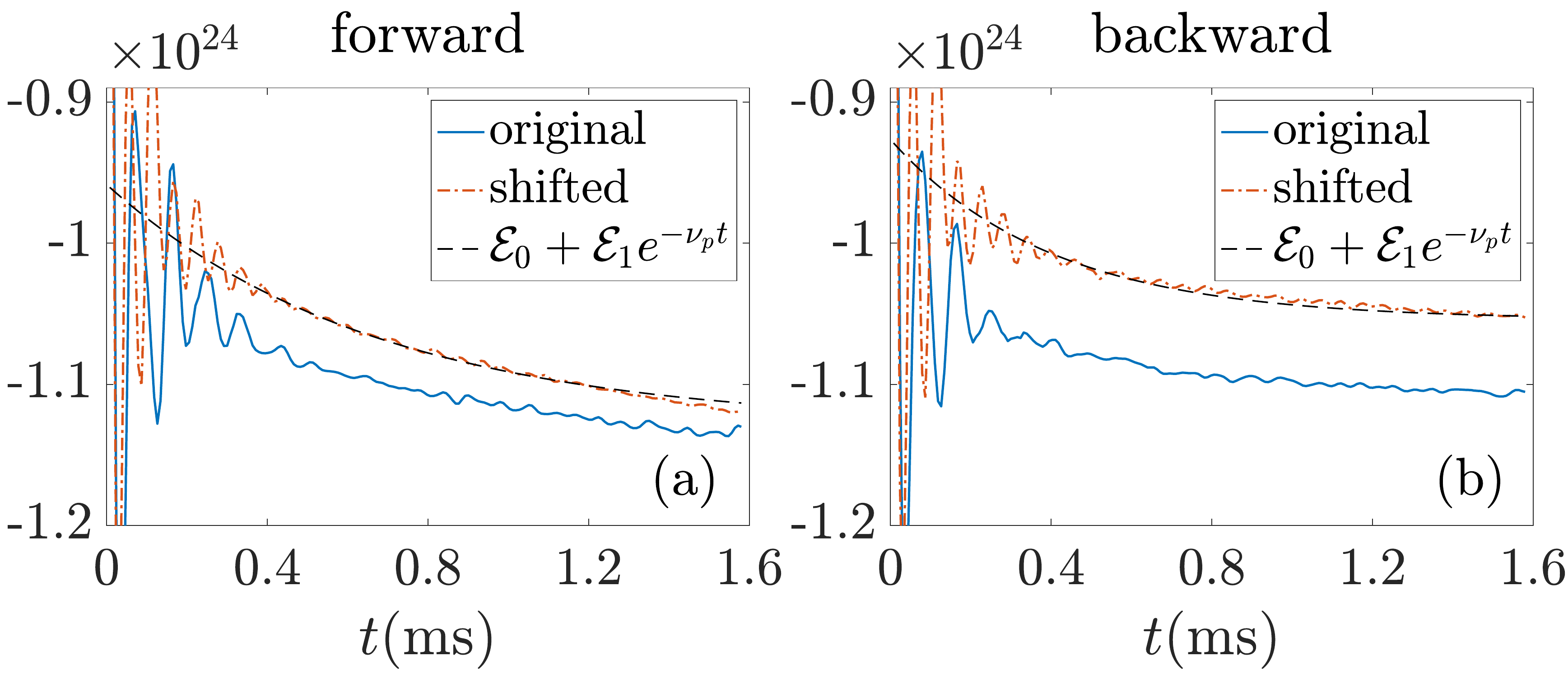}\caption{Comparisons of $\mc{E}$ (in units ${\rm eV}/({\rm T}\cdot{\rm m}^5)$) between simulations using the original equilibrium profile  (blue solid curves) and the radially shifted profile (red dot-dashed curves). For the original profile, $\mc{E}$ is evaluated at $\psi_n=0.984$; and for the shifted profile, $\mc{E}$ is evaluated at $\psi_n=0.884$.  To reduce numerical fluctuations, the data has been smoothed by averaging over a time window of $\Delta t=0.04$ms. Also shown are fitting curves $\mc{E}=\mc{E}_0+\mc{E}_1 e^{-\nu_p t}$ that model the collisional relaxation. In (a), $\mc{E}_0=-1.13\times 10^{24}$, $\mc{E}_1=1.71\times 10^{23}$, and $\nu_p=1.5\times 10^3{\rm s}^{-1}$. In (b), $\mc{E}_0=-1.05\times 10^{24}$, $\mc{E}_1=1.28\times 10^{23}$, and $\nu_p=2.5\times 10^3{\rm s}^{-1}$. }\label{shiftEr}
\end{figure}

To estimate $\nu_p$ without the effects from ion orbit loss, we ran XGCa simulations again, but with the equilibrium profile of $n_{i0}$ and $T_{i0}$ radially shifted inwards by $\Delta\psi_n=0.1$. \Fref{shiftEr} shows a comparison of $E_r$ between the simulations with the original profile and the shifted profile. The magnitude of $|E_r|$ is smaller by a few percent for the shifted equilibrium profile. However, the collisional damping rate of $E_r$ is similar between the two profiles, suggesting that the collisional damping mechanism is similar whether there is gyrocenter orbit loss or not.  By fitting the numerical results to an analytic form $\mc{E}=\mc{E}_0+\mc{E}_1e^{-\nu_p t}$, we estimate $\nu_p\approx 1.5\times 10^{3}{\rm s}^{-1}$ for the forward-$\nabla B$ case, and $\nu_p\approx 2.5\times 10^{3}{\rm s}^{-1}$ for the backward-$\nabla B$ case (\fref{shiftEr}). Both estimates of $\nu_p$ are the same order of magnitude as the local ion-ion collision rate $\nu_{i}\approx 1.6\times 10^{3}{\rm s}^{-1}$, which is calculated 
using $n_{i0}\approx 10^{19}{\rm m}^{-3}$ and $T_{i0}\approx 200{\rm eV}$ at the edge \cite{Huba98}.
Assuming $\nu_p= 1.5\times 10^{3}{\rm s}^{-1}$, together with $\Gamma_r^{\rm loss}\approx 10^{19}{\rm s}^{-1}$ and $\int\hat{J}_0\delta n_i\,dV\approx 3\times 10^{17}$, we estimate the relative change of $E_r$ as
\begin{equation}
\left|\frac{\Delta E_r}{E_r}\right|\approx\left|\frac{\Gamma_r^{\rm loss}}{\nu_p\int \hat{J}_0\delta n_i\, dV}\right|\approx 2.2\%.
\end{equation}
This is consistent with \fref{shiftEr}, in which $\mc{E}$ for the shifted profiles is within a few percent of $\mc{E}$ for the original profiles, although the shifted-profile calculations involve almost no loss orbits. For these simulations, we thus conclude that $\Gamma_r^{\rm loss}$ is too small to drive $E_r$ significantly away from $E_r^{\rm neo(0)}$. A modest difference in $E_r$ is observed between the forward- and backward-$\nabla B$ configurations. However, a similar difference also exists in the shifted-profile simulations (\fref{shiftEr}), suggesting that it may in fact follow from the difference in the toroidal rotation profile.
\subsection{Toroidal-rotation acceleration at the edge}
\label{edge_rotation}
\begin{figure}
\includegraphics[width=1\columnwidth]{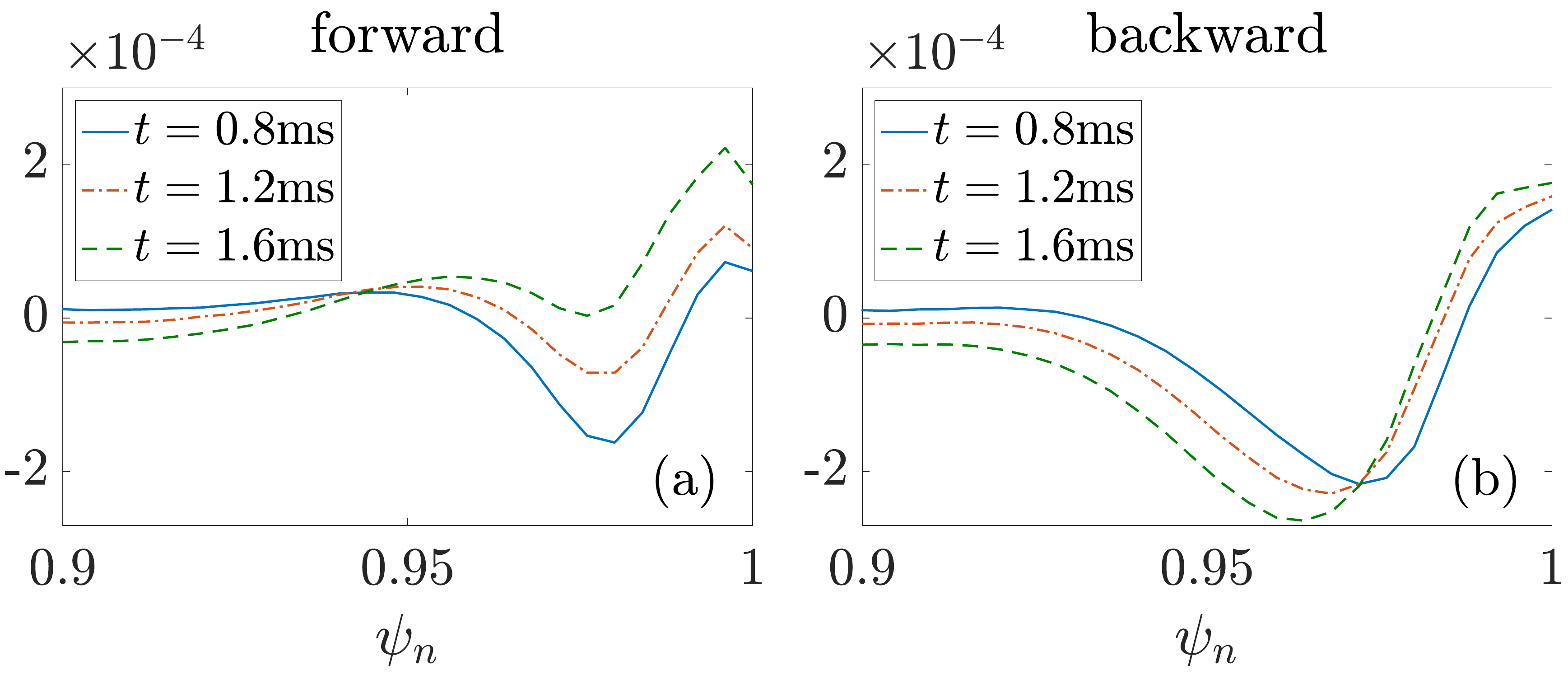}\caption{Time evolution of the flux-surface averaged toroidal angular momentum density (in units ${\rm kg}/({\rm m}\cdot{\rm s})$) at the edge. }\label{edgeRot}
\end{figure}

The fact that $\Gamma_t$ is nonzero indicates that the gyrocenter distribution function $F_i$ is changing over time. Since the density moment of $F_i$  is not changing much, consistent with the quasisteady $E_r$ profile, we will look for a change in the first moment of $F_i$, namely, the parallel flow velocity. \Fref{edgeRot} shows the flux-surface averaged parallel-flow contribution to the toroidal angular momentum density $\avg{\int d\mc{W}F_ip_\parallel R B_\varphi/B}$ in the edge. Because $B_\theta>0$,  the equilibrium plasma current is oppositely directed to $\nabla\varphi$; hence, a positive (negative) toroidal angular momentum corresponds to a counter- (co-) current rotation. As shown in the figure,  at an earlier time the toroidal rotation has a co-current peak inside the LCFS,  consistent with previous results \cite{Chang04,Seo14,Chang08}. As time progresses, however, there is a counter-current  acceleration just inside the LCFS for both configurations. (As one moves further inside radially, the acceleration shifts to the co-current direction.) Both the toroidal-rotation velocity and acceleration are much stronger at the outboard than at the inboard. 

The counter-current toroidal acceleration just inside the LCFS is consistent with the negative $\Gamma_t$.  Counter-current acceleration at fixed density implies an increase of counter-current ions and a decrease of co-current ions.  At the outboard midplane, where the acceleration is concentrated, most co-current ions are on orbits that remain inside the LCFS, which do not contribute to the orbit flux.  Many counter-current ions are on orbits that do cross the LCFS, thus the counter-current acceleration results in a net positive $\pd_t F_i$ in the counter-current velocity space in \eref{formulation_final}, which corresponds to a negative $\Gamma_t$.

For electrostatic simulations with the adiabatic-electron model \eref{XGC_adia_elec}, the toroidal angular momentum is conserved  \cite{Scott10,Stoltzfus17}. Although the toroidal angular momentum consists of both a parallel-momentum portion (shown in \fref{edgeRot}) and an $\ExB$ portion, the latter is almost constant in time here, because $E_r$ is quasisteady. The observed toroidal-rotation acceleration therefore indicates a change of the toroidal-angular-momentum density, hence a nonzero radial toroidal-angular-momentum flux.
We also observed qualitatively similar counter-current toroidal-rotation acceleration in simulations with shifted equilibrium profiles, suggesting that this acceleration is not solely driven by effects specific to the edge. These observations will be the subject of future studies.
\section{Conclusions and discussion}
\label{conclusion}
The ion-orbit-flux formulation \cite{Stoltzfus20,Stoltzfus21} has been implemented as a numerical diagnostic in XGCa \cite{XGC,ion-orbit,orbit-flux}. The diagnostic measures the separate contributions to the ion orbit loss from different transport mechanisms and sources. The validity of the diagnostic is demonstrated by studying the collisional relaxation of $E_r$ in the core. Then, the diagnostic is used to study effects of ion orbit loss on $E_r$ at the edge of a DIII-D H-mode plasma. Under the given neoclassical pedestal plasma density and temperature profiles and without considering neutral particles, the radial electric force on ions from $E_r$ approximately balances the ion radial pressure gradient in the edge pedestal. The existence of a small ion gyrocenter orbit loss flux does not drive $E_r$ significantly away from its standard neoclassical solution, because of the collisional poloidal-rotation damping and the large radial dielectric response of tokamak plasma. In other words, once the edge pedestal is established by some mechanism, its $E_r$ is not very sensitive to the existence of a small radial current perturbation.

The role of collisional ion orbit loss in a full-current ITER edge plasma is important and is left for a subsequent study. A recent report has shown that due to weak neoclassical effect in the full-current ITER, the plasma pressure gradient near the magnetic separatrix will not be balanced by $E_r$, but will be balanced by strong toroidal rotation driven by turbulent orbit loss \cite{Chang21}. It is of interest to see if similar results also hold for an axisymmetric neoclassical ITER plasma.
\ack
This work was supported by the U.S. Department of Energy through Contract No. DE-AC02–09CH11466. Funding to R. Hager, S. Ku and C.S. Chang is provided via the SciDAC-4 program.
The simulations presented in this article were performed on computational resources managed and supported by Princeton Research Computing, a consortium of groups including the Princeton Institute for Computational Science and Engineering (PICSciE) and the Office of Information Technology's High Performance Computing Center and Visualization Laboratory at Princeton University.  
This research used resources of the National Energy Research Scientific Computing Center, which is supported by the Office of Science of the U.S. Department of Energy under Contract No. DE-AC02-05CH11231. 
\section*{Data Availability}
Digital data can be found in DataSpace of Princeton University \cite{data}.
\section*{References}
\bibliographystyle{iopart-num}
\bibliography{references}

\end{document}